\documentclass[a4paper,12pt]{article}
\usepackage{amssymb}
\usepackage[latin1]{inputenc}
\begin{document}
\def\Z{\sf Z\hspace{-1.8mm}Z\hspace{.1em}}
\newcommand{\farc}[2]{\;\; \frac{\displaystyle{#1}}{\displaystyle{#2}}\;\; }
\def\pe{\hbox{\large \bf [}}
\def\pd{\hbox{\large\bf ]}}
\def\pet{\hbox{\large\bf [}}
\def\pdt{\hbox{\large\bf ]}}
\def\bet{\hbox{\large (}}
\def\bdt{\hbox{\large )}}
\setcounter{secnumdepth}{8}
\setlength{\parindent}{1cm} 
\setlength{\baselineskip}{20pt}
\setlength{\normalbaselineskip}{20pt}
\addtolength{\parskip}{3pt}
\setlength{\abovedisplayskip}{16pt plus3pt minus7pt}
\setlength{\belowdisplayskip}{\abovedisplayskip}
\setlength{\abovedisplayshortskip}{12pt plus3pt}
\setlength{\belowdisplayshortskip}{\abovedisplayshortskip}
\newcommand{\spar}{\hspace{\parindent}}
\renewcommand{\thesubsection}{\thesection\arabic{subsection}.}
\renewcommand{\thesubsubsection}{\thesubsection\arabic{subsubsection}.}
\def\diss{\rm diss}

\setcounter{page}{1}

\begin{center}
{\bf ANÁLISE TERMODINÂMICA DE UM CIRCUITO RLC}
\end{center}

\begin{center}
Rodrigo de Abreu \\
Centro de Electrodinâmica e Departamento de Física do IST 
\end{center}

\noindent
\underline{\bf Abstract} \\ 
We analyse a RLC circuit taking the Second Law of Thermodynamics into \\ consideration. \\

\noindent
\underline {\bf Introdução}

Hoje em dia, torna-se cada vez mais difícil definir onde se situa a 
fronteira entre a Mecânica e a Termodinâmica. Em primeiro lugar, torna-se 
difícil em inúmeros problemas fazer uma distinção clara, e com significado 
físico, entre os conceitos de trabalho e calor associados à Primeira Lei da 
Termodinâmica [1,2]. Em segundo lugar, como é sabido, uma das vias para a 
interpretação física de alguns conceitos termodinâmicos, como o calor 
e a entropia, é a partir da Mecânica Estatística.
Por último, na maior parte dos problemas de 
Mecânica envolvendo sistemas dissipativos, a análise realizada debruça-se 
exclusivamente sobre os aspectos mecânicos, não sendo feita qualquer 
tentativa para uma abordagem termodinâmica do problema.

O problema que iremos aqui abordar é, na sua essência, extremamente simples,
sendo descrito pela Lei de Hooke. Esta lei 
postula a proporcionalidade entre a força aplicada a uma mola
e o elongamento sofrido por esta
\begin{equation}
F = k \; (L_1 - L_0) \; ,
\end{equation}
onde $k$ é a constante de restituição da mola e $L_0$ e $L_1$ representam,
respectivamente, os comprimentos inicial e final da mola. Uma mola que
obedeça a (1), com $k$ constante em qualquer circunstância, é chamada aqui 
{\it mola ideal}. Uma outra situação mais real, onde a proporcionalidade 
anterior já não se verifique, irá ser tratada igualmente neste trabalho numa
secção à parte.   

\noindent
{\bf 1 $-$} \underline{\bf A Mola Ideal como Sistema Dissipativo}

Considere-se então em primeiro lugar uma mola ideal, de constante $k$ e 
comprimento $L_0$, à qual se aplica uma força constante $F_0$, por exemplo, 
pendurando-se na vertical uma massa de peso $F_0$=$m g$. Como é 
sabido, a mola sofre uma elongação até à sua nova posição de equilíbrio 
$L_1$, mesmo que a mola se encontre no vácuo, sendo este ponto simplesmente 
determinado por
\begin{equation}
k \; (L_1 - L_0) = F_0 \; .
\end{equation}
O trabalho realizado pela força aplicada exterior na elongação da mola
é usando (2) dado por
\begin{equation}
\tau = \int_{L_0}^{L_1} F_0 \, dx = F_0 \; (L_1 - L_0) = \frac{F_0^{\;2}}{k} 
\; ,
\end{equation}
enquanto que a energia potencial por esta adquirida é dada por
\begin{equation}
V = \int_{L_0}^{L_1} k \; x \; dx = \frac{1}{2} \; k \; (L_1 - L_0)^2 = 
\frac{1}{2} \; \frac{F_0^{\;2}}{k} \; ,
\end{equation}
isto é, igual a metade do trabalho $\tau$. Desta diferença 
emerge imediatamente a seguinte pergunta: Para onde vai então a parte 
restante do trabalho realizado pela força $F_0$? Não fica concerteza
armazenado no sistema
sob a forma de energia cinética, dado que a mola fica parada no
fim do movimento.

A resposta a esta questão, reside obviamente no facto de termos considerado 
a elongação da mola como um processo sem atrito. Na realidade, a elongação 
da mola ocorre sempre com dissipação de energia e quaisquer que sejam as 
características da mola ou do meio circundante que a envolva, a energia 
dissipada é exactamente igual a metade do trabalho realizado pela força 
aplicada $F_0$. Isto é, mesmo que o meio circundante seja o vácuo, onde 
não existe atrito com o ar, a energia dissipada apresenta sempre o mesmo 
valor, $W_{diss.}$=$\tau$/2.

Uma situação de certo modo análoga surge quando se carrega um condensador
de capacidade $C$, inicialmente descarregado, ligando-o a uma bateria de 
força electromotriz $\epsilon$. Vamos supor que a carga do 
condensador é feita ligando-se em série, com o condensador e a bateria, 
uma resistência $R$, a qual inclui a resistência da própria bateria. A 
solução deste problema é bem conhecida dos cursos de Electromagnetismo. A 
diferença de potencial ({\it d.d.p.}) aos terminais do condensador é dada 
por
\begin{equation}
v_c(t) = \epsilon \; (1 - e^{-t/RC}) \; ,
\end{equation}
pelo que o condensador vai acabar por ficar carregado com uma {\it d.d.p.}
$v_c(t)$=$\epsilon$, quando $t$$\to$$\infty$, e portanto com uma energia
electrostática final igual a
\begin{equation}
W_e = \frac{1}{2} \; C \; \epsilon^2 \; .
\end{equation}
Por outro lado, a corrente que percorre o circuito, enquanto ele estiver
a ser carregado, é dada por
\begin{equation}
i= \frac{\epsilon}{R}  \; e^{-t/RC} \; ,
\end{equation}
o que permite calcular imediatamente, tanto a energia dissipada por efeito
de Joule na resistência $R$
\begin{equation}
W_{diss.} = \int_0^{\infty} R \; i^2 \, dt = \frac{1}{2} \; C \; 
\epsilon^2 \;,
\end{equation}
como o trabalho fornecido pela bateria
\begin{equation}
\tau  = \int_0^{\infty} \epsilon \; i \, dt =  C \; \epsilon^2 \;.
\end{equation}
Tem-se aqui também, tal como no caso da mola, que a energia dissipada na 
resistência é exactamente igual a metade do trabalho fornecido pela bateria; 
e este resultado é tanto mais notável quando se observa que o resultado a 
que se chega é independente do valor da resistência $R$. Isto é, se fizermos 
tender $R$$\to$0, o condensador é carregado num tempo extremamente curto, 
com um valor inicial de corrente $i(0)$$\sim$$\infty$. Contudo, a energia 
total dissipada mantem-se sempre constante, apresentando o valor 
$W_{diss.}$=$\tau$/2.

Ora, a situação que aqui pretendemos tratar, de uma mola ideal sujeita
a uma força constante $F_0$, apresenta semelhanças óbvias com o caso do 
condensador. Em primeiro lugar, em vez de nos preocuparmos em determinar 
apenas o ponto de equilíbrio dado por (2), temos de considerar que a 
equação que descreve a elongação da mola é na realidade dada por 
\begin{equation}
m \; \frac{d^2x}{dt^2} + \beta \; \frac{dx}{dt} + k \; x = F_0 \; ,
\end{equation}
sendo $x$=($L$$-$$L_0$) o valor da elongação da mola em cada instante e onde
fenomenologicamente se introduziu uma força de atrito $F_{at.}$=$\beta$$v$, 
tendo em conta os efeitos dissipativos que ocorrem na mola. A equação (10)
apresenta como solução, no caso de um regime oscilante amortecido 
\begin{equation}
x(t) = \frac{F_0}{k} \; + \; A  \; e^{-\lambda t} \cos(\omega t + \phi) \; ,
\end{equation}
com $\lambda$=$\beta$/$2m$ e $\omega$=$\sqrt{(k/m)-\lambda^2}$. No caso da
mola se encontrar parada no início, as condições de continuidade em $t$=0
para as energias potencial e cinética, determinam as
constantes de integração: $A$=$-$($F_0$/$k$)$\; \sec \phi$; 
$\phi$=$-$$\arctan (\lambda/\omega)$. No limite $t$$\to$$\infty$, tem-se 
como anteriormente ($L_1$$-$$L_0$)=$F_0$/$k$. 

Os sistemas mecânico e eléctrico aqui descritos apresentam efectivamente o
mesmo tipo de soluções. A aparente diferença entre as soluções (5) e (11)
resulta simplesmente do facto de termos considerado nulo o coeficiente 
de indução do circuito. Se ao invés, tivessemos admitido a existência de um
elemento indutivo no circuito, as soluções seriam também do tipo oscilante
amortecido. Contudo, o balanço energético manter-se-ia inalterado, na
medida em que o elemento indutivo não dissipa nem armazena energia no final.
Note-se que após o condensador estar carregado $i$=0, pelo que a energia
magnética no elemento indutivo é também nula.

Se multiplicarmos agora ambos os membros de (10) pela variação elementar
$dx$ e integrarmos essa equação entre $L_0$ e $L_1$, obtemos a equação
que traduz o balanço de energia na mola. O primeiro termo do membro de lado
esquerdo é obviamente nulo, na medida que não existe variação de energia 
cinética entre os pontos de equilíbrio inicial e final, o segundo termo
representa a energia dissipada na mola $W_{diss.}$, o terceiro termo
representa a variação de energia potencial dada por (4), enquanto que
o termo do lado direito é o trabalho (3) realizado pela força exterior.
Este simples balanço permite-nos escrever para a energia dissipada 
no termo de atrito $\beta$
\begin{equation}
W_{diss.} = \int_{L0}^{L1} 
 \beta \; \frac{dx}{dt} \; dx = \frac{1}{2} \; F_0 \; (L_1-L_0) \;.
\end{equation}

É interessante notar ainda que a conclusão a que se chegou é independente
do valor de $\beta$, podendo inclusivamente ter-se $\beta$$\to$0. Este caso
corresponde no circuito $RC$ à situação $R$$\to$0, em que 
$i(0)$$\sim$$\infty$, mas onde a energia dissipada por efeito de Joule 
continua, mesmo neste limite, a ser dada por (8). Regressando agora ao
caso da mola e supondo $m$$\to$0, de forma ao sistema ser descrito
igualmente
por uma equação diferencial de primeira ordem (correspondente ao limite
$L$$\to$0 no circuito $RLC$), ao limite 
$\beta$$\to$0 corresponde uma velocidade inicial infinita 
($dx$/$dt$)$_0$$\sim$$\infty$. Contudo, a energia dissipada na mola continua
a ser dada por (12). Assim, embora este facto não seja normalmente tratado 
nos cursos elementares, uma mola ideal, sujeita a uma força aplicada
$F_0$ constante, é sempre dissipativa. Este facto
verifica-se independentemente das características materiais da mola. 

\noindent
{\bf 2 $-$} \underline{\bf Análise Termodinâmica da Mola Ideal}

\noindent
O resultado a que se chegou em (12) mostra que a energia dissipada na mola
é sempre igual a metade do trabalho realizado pela força aplicada exterior,
independentemente do valor do termo de atrito $\beta$. Assim, mesmo que 
o meio onde a mola se encontre seja o vácuo e que não exista atrito com 
o ar, há sempre um atrito interno da própria mola. A energia dissipada 
internamente dá origem a uma elevação de temperatura da mola acompanhada 
por uma transferência de energia para o ambiente, esta última sob a forma 
de radiamento no caso do meio envolvente ser o vácuo.

Este problema pode ser tratado agora sob o ponto de vista termodinâmico.
Em primeiro lugar, a entropia da mola obedece à equação de estado
$S$=$S$($T$,$L$), pelo que uma variação elementar desta é dada por
\begin{equation}
dS = \left(\frac{\partial S}{\partial T} \right)_L dT +
\left(\frac{\partial S}{\partial L} \right)_T dL \;,
\end{equation}
enquanto que para a energia interna da mola podemos escrever a relação 
termodinâmica fundamental para um processo infinitesimal
\begin{equation}
dU = T \; dS + F \; dL \;. 
\end{equation}
Como sabemos, a capacidade calorífica para um comprimento $L$ é dada por
\begin{equation}
C_L = \left(\frac{\partial U}{\partial T} \right)_L =
T \left(\frac{\partial S}{\partial T} \right)_L \;,
\end{equation}
o que nos permite obter para o primeiro termo de (13)
\begin{equation}
\left(\frac{\partial S}{\partial T} \right)_L = 
\frac{C_L}{T} \;.
\end{equation}
Por outro lado, o segundo termo de (13) pode ser calculado usando uma das 
relações termodinâmicas de Maxwell, com as variáveis habituais pressão e 
volume substituídas aqui por $F$ e $L$, pelo que usando (1), onde $k$ é 
independente de $T$, podemos escrever
\begin{equation}
\left(\frac{\partial S}{\partial L} \right)_T =
- \left(\frac{\partial F}{\partial T} \right)_L = 0 \; .
\end{equation}

Temos assim que neste caso a entropia pode ser calculada a partir da relação
\begin{equation}
dS = \frac{C_L}{T} \; dT \; ,
\end{equation}
a qual uma vez integrada, admitindo que $C_L$ não depende de $T$, permite 
obter para a variação de entropia, entre os estados inicial 
$S$=$S$($T_0$,$L_0$) e final $S$=$S$($T_1$,$L_1$), a seguinte expressão 
\begin{equation}
\Delta S = C_L \; \ln \left ( \frac{T_1}{T_0} \right ) \;.
\end{equation}
Neste caso, a transformação é isentrópica se $T_1$=$T_0$.

No que se refere agora à energia interna, usando (1) e (18) podemos escrever 
(14) sob a forma
\begin{equation}
dU = C_L \; dT + k \; (L-L_0) \; dL \;, 
\end{equation}
obtendo para a variação de energia interna entre os mesmos estados inicial
e final, 
\begin{equation}
\Delta U = C_L \; (T_1-T_0) + \frac{1}{2} \; k \; (L_1-L_0)^2 \;.
\end{equation}
 
Se admitirmos que não há transferência de energia para o ambiente, a
variação de energia interna da mola $\Delta U$ é igual ao trabalho realizado
pela força aplicada exterior na elongação da mola (3)
\begin{equation}
\tau = \frac{F_0^2}{k} \;,
\end{equation}
sendo o comprimento final determinado por (2), ($L_1$$-$$L_0$)=$F_0$/$k$.
O aumento de temperatura é dado assim por
\begin{equation}
T_1 = T_0 + \frac{1}{2} \; \frac{F_0^{\;2}}{k \; C_L} \;,
\end{equation}
enquanto que para a variação de entropia tem-se
\begin{equation}
\Delta S =C_L \; \ln \left ( 1 + \frac{1}{2} \; 
\frac{F_0^{\;2}}{k \; C_L \; T_0} \right ) \;.
\end{equation}
Esta variação de entropia foi calculada admitindo que a mola está 
termicamente isolada do ambiente, pelo que é devida a um fenómeno de atrito 
interno resultante da própria deformação do material.

Se ao contrário admitirmos agora que a mola é um sistema não isolado,
de forma a que os valores de temperatura e entropia possam regressar aos
seus valores iniciais por troca com o exterior, a energia dissipada na 
mola acaba por ser transferida integralmente para o ambiente, produzindo 
neste um aumento $\Delta U_0$=$W_{diss.}$. Podemos neste caso estimar o 
aumento de entropia do ambiente, considerando que este se comporta como
uma "fonte de calor" ($T_0$=$const.$), usando
(12) e (14)
\begin{equation}
\Delta S_0 = \frac{\Delta U_0}{T_0} = \frac{1}{2} \; 
\frac{F_0^{\;2}}{k \; T_0} \;.
\end{equation}

\noindent
{\bf 3 $-$} \underline{\bf Entropia e Energia Interna de uma Mola Não
Ideal}

\noindent
A análise anterior torna-se um pouco mais complicada no caso de 
considerarmos uma mola não ideal. Assim, considere-se o exemplo referido em 
[3], onde é suposto que num dado intervalo de temperaturas, o comprimento 
$L$ de uma barra, designada aqui em sentido lato por mola, está relacionado 
com a força $F$ que lhe é aplicada através da expressão
\begin{equation}
F = a \; T^2 \; (L - L_0) \; ,
\end{equation}
onde $a$ é uma constante positiva, $L_0$ é o comprimento da barra quando 
não sujeita a qualquer força e $T$ é a temperatura da barra quando lhe é
aplicada a força $F$. Esta barra comporta-se como uma mola não ideal,
na medida em que neste caso 
o termo que correspondia anteriormente à constante de restituição,
$k(T)$=$a T^2$, não é agora constante.
Note-se que como iremos mostrar a seguir, mesmo quando a temperatura se 
mantem constante, e igual à temperatura ambiente $T_0$, a mola não se
comporta como ideal.

Comecemos por determinar as variações de entropia $S$ e energia interna $U$
quando a barra é alongada de $L_0$ para $L_1$ e a temperatura varia entre
$T_0$ e $T_1$. Tal como em [3], admitamos que a capacidade calorífica para
o comprimento $L_0$ é dada por $C_{L_0}$=$b T$, sendo $b$ uma constante
positiva. Para qualquer outro comprimento tem-se como habitualmente $C_L$ 
dado por (15). Neste caso, o segundo termo de (13), obtido a partir de uma 
das relações de Maxwell, toma a forma
\begin{equation}
\left(\frac{\partial S}{\partial L} \right)_T = - 
\left(\frac{\partial F}{\partial T} \right)_L = - 2 \; a \; T \; (L-L_0) \;.
\end{equation}

Calculemos a variação de entropia $\Delta$$S$ supondo que a passagem do 
estado inicial ($L_0$,$T_0$) para o estado final ($L_1$,$T_1$), ocorre 
através de um estado intermédio ($L_0$,$T_1$). De (13), (16) e (27), 
podemos escrever 
\begin{equation}
(dS)_{L_0} = \frac{C_{L_0}}{T} \; dT = b \; dT 
\end{equation}
e
\begin{equation}
(dS)_{T_1}= -2 \; a \; T_1 \; (L-L_0) \; dL \;.
\end{equation}
Integrando agora (28) e (29), obtém-se para a variação de entropia entre os 
estados inicial e final
\begin{equation}
\Delta S = b \; (T_1 - T_0) - a \; T_1 \; (L_1 - L_0)^2 \;.
\end{equation}

No que se refere à variação de energia interna $\Delta$$U$, ela pode ser 
calculada diferenciando (30)
\begin{equation}
dS = b \; dT - a \; (L - L_0)^2 \; dT - 2 \; a \; T \; (L - L_0) \; dL 
\end{equation}
e substituindo esta expressão em (14) usando (26)
\begin{equation}
dU = b \; T \; dT  -  a \; T \; (L - L_0)^2 \; dT - a \; T^2 
\; (L - L_0) \; dL \;.
\end{equation}
Uma vez integrada esta última expressão, podemos escrever
\begin{equation}
\Delta U = \frac{1}{2} \; b \; (T_1^{\;2} - T_0^{\;2}) 
- \frac{1}{2} \; a \; T_1^{\;2} \; (L_1 - L_0)^2 \;.
\end{equation}

É interessante analisarmos antes de mais a que nos conduzem estes resultados
numa transformação em que a temperatura se mantenha sempre constante.
Assim, vamos supor que a elongação da mola se faz progressivamente, por 
acréscimos infinitesimais da força exterior, suficientemente lentos para 
que a mola se mantenha à temperatura ambiente $T_0$. Trata-se pois de uma 
situação onde, de acordo com (30) e (33), as variações de entropia e de 
energia da mola são dadas por 
\begin{equation}
\Delta S = - a \; T_0 \; (L_1 - L_0)^2 \;;  
\end{equation}
\begin{equation}
\Delta U = - \frac{1}{2} \; a \; T_0^{\;2} \; (L_1 - L_0)^2 \;.
\end{equation}
No caso concreto que estamos aqui a tratar, a transformação é reversível, 
dado que ocorre através de uma sucessão de estados de equilíbrio. Nestas 
condições, a transformação é isentrópica para o sistema global, constituído 
pela mola e o ambiente que a rodeia ($\Delta S$+$\Delta S_0$=0), podendo 
portanto escrever-se de (34)
\begin{equation}
\Delta S_0 = - \Delta S = a \; T_0 \; (L_1 - L_0)^2 \;.
\end{equation}

Por outro lado, o trabalho realizado pela força aplicada (26) na elongação 
da mola, neste caso em que a transformação é reversível, com $T$=$T_0$ ao 
longo da transformação, é dado por
\begin{equation}
\tau = \int^{L_1}_{L_0} a \; T_0^{\;2} \; (L - L_0) \; dL = 
 \frac{1}{2} \; a \; T_0^{\;2} \; (L_1 - L_0)^2 \;.
\end{equation}
Nestas condições a variação de energia interna $\Delta U$ é dada por
\begin{equation}
\Delta U = - \Delta U_0 + \tau \; ,
\end{equation}
sendo $\Delta U_0$ a energia transferida para o ambiente, a qual é neste 
caso igual a duas vezes o trabalho realizado pela força exterior na 
elongação da mola
\begin{equation}
\Delta U_0 = 2 \; \tau = a \; T_0^{\;2} \; (L_1 - L_0)^2 \;,
\end{equation}
verificando-se ainda $\Delta U_0$=$T_0$ $\Delta S_0$. 

Repare-se que o resultado a que acabámos de chegar é assaz curioso. Não só 
o trabalho realizado na elongação da mola é integralmente transferido
para o ambiente, como também a energia interna é reduzida exactamente da
mesma quantidade, a qual é também transferida para o ambiente. Se 
procurassemos agora analogias com sistemas eléctricos, encontrariamos
uma situação em tudo semelhante, quando se afasta entre si duas armaduras
de um condensador plano a potenciais constantes. Neste caso, o trabalho
realizado pela força exterior necessário ao afastamento das armaduras
é integralmente enviado para a bateria, assim como a própria energia 
electrostática do sistema diminui do mesmo valor sendo essa energia
transferida também para a bateria. A bateria comporta-se aqui como o
ambiente no caso da mola, recebendo tanto o trabalho realizado pela força
exterior, como a diminuição da energia electrostática do condensador.
Obviamente o resultado a que se chegou depende da lei de variação com a
temperatura (26). Se outra dependência existisse já esta analogia não se
verificaria.

O caso que acabamos de estudar corresponde à situação limite de uma
transformação infinitamente lenta e onde a mola se comporta como um 
sistema não isolado. Considere-se agora o caso em que a mola se encontra 
isolada termicamente do ambiente e onde a sua elongação ocorre de uma forma 
isentrópica, $\Delta$$S$=0. De (30), podemos escrever
\begin{equation}
b \; (T_1 - T_0) = a \; T_1 \; (L_1 - L_0)^2 
\end{equation}
e portanto
\begin{equation}
T_1 = \frac{b \; T_0}{b - a \; (L_1-L_0)^2} \;.
\end{equation}
Existe agora um aumento da temperatura com a elongação da mola; 
$T_1$$>$$T_0$ para $L_1$$\neq$$L_0$. Substituindo (41) em (33), obtém-se 
para a variação de energia interna da mola,
\begin{equation}
\Delta U = \frac{1}{2} \; a \; \frac{(L_1-L_0)^2}
{b-a \; (L_1-L_0)^2} \; b \; T_0^{\;2} \;,
\end{equation}
e tendo de novo em conta (41),
\begin{equation}
\Delta U = \frac{1}{2} \; a \; T_1 \; T_0 \; (L_1-L_0)^2 \;.
\end{equation}

Entre dois pontos em que a temperature não varie apreciavelmente, tem-se
\begin{equation}
\Delta U = \frac{1}{2}\; k^{\prime} \; (L_1-L_0)^2 \;, 
\end{equation}
com $k^{\prime}$=$aT_1T_0$$\simeq$$const.$ Aparentemente encontramos de novo a
situação da mola ideal, pois entre dois pontos próximos $T_1$$\simeq$$T_0$ e
$k^{\prime}$ toma um valor próximo da constante de restituição da mola 
$k$=$a$$T^2$. Contudo, o primeiro termo de (21) é agora desprezável, pelo 
que a energia interna da mola só possui o termo de energia potencial.

A situação descrita por (43) tem algum interesse, pois permite aparentemente 
reeencontrar a expressão vulgarmente usada para a energia de uma mola.
Contudo, no caso presente, (43) não representa uma energia potencial.

\noindent
{\bf 4 $-$} \underline{\bf Conclusões}

\noindent
A Mecânica e a Termodinâmica são habitualmente tratadas como dois capítulos
praticamente estanques da Física, tornando-se muito difícil a interpretação 
de certos fenómenos à luz dos dois conceitos. Neste trabalho procurou-se 
contribuir, usando um sistema extremamente simples, para que esta 
integração possa vir para a ordem do dia, chamando a atenção para o
facto de que quando se tratam sistemas dissipativos, é desejável que uma
análise termodinâmica seja feita também em paralelo com o estudo mecânico.
Só desta forma podemos entender o que acontece, de facto, a grandezas
tão importantes como a energia interna ou a entropia de um sistema.

Ao longo deste trabalho começou-se por mostrar que a energia dissipada,
na elongação de uma mola ideal, por aplicação de uma força 
$F_0$ constante, é sempre igual a metade do trabalho
necessário para o fazer, independentemente do meio envolvente. Se este for 
o vácuo, a energia é dissipada internamente na própria mola, produzindo 
uma elevação de temperatura, a qual permanecerá ou não indefinidamente 
na mola consoante esta estiver ou não isolada do ambiente. No caso de 
uma mola ideal, foram calculados os aumentos de entropia da mola e do 
ambiente, nos casos em que a mola está, respectivamente, isolada e não 
isolada termicamente do exterior.

Por último, foi considerado ainda o caso de uma mola não ideal, em que a
constante de restituição da mola é uma função da temperatura. Aqui duas
situações distintas foram analisadas: Sistema não isolado com a elongação
da mola a ter lugar através de um processo infinitamente lento; Sistema
isolado com a elongação da mola a ocorrer de uma forma isentrópica. Os
aspectos mais relevantes para cada situação foram devidamente evidenciados
e discutidos. \\  

\noindent
$[$1$]$ $-$ Laufer G., Am. J. Phys. \underline{51}, (1983) 42. \\
$[$2$]$ $-$ Barrow G. M., J. Chem. Educ. \underline{65}, (1988) 122. \\
$[$3$]$ $-$ Reif F., Fundamentals of Statistical and Thermal Physics 
(McGraw-Hill, Tokyo, 1965), p.196. \\

\noindent
Agradeço ao Prof. Jorge Loureiro do IST todas as críticas, sugestões
e revisão cuidadosa deste manuscrito.

\end{document}